\documentclass{optica-article}
\journal{opticajournal} 

\articletype{Research Article}

\usepackage{lineno}

\usepackage{amsmath,amsfonts}%
\usepackage{mathrsfs}%
\usepackage{textcomp}%
\usepackage{manyfoot}%
\usepackage{booktabs}%
\usepackage{algorithm}%
\usepackage{algorithmicx}%
\usepackage{algpseudocode}%
\usepackage{listings}%

\usepackage{xcolor} 

\renewcommand{\thesubsection}{\Alph{subsection}}

\usepackage{bibunits}

\newcommand{\compilemode}{both}

\begin{document}

\def\doMain{0}
\def\doSupp{0}
\edef\temp{\compilemode}
\def\mainStr{main}
\def\suppStr{supp}
\def\bothStr{both}
\ifx\temp\mainStr
  \def\doMain{1}\def\doSupp{0}
\else\ifx\temp\suppStr
  \def\doMain{0}\def\doSupp{1}
\else
  \def\doMain{1}\def\doSupp{1}
\fi\fi

\ifnum\doMain=1
  \begin{bibunit}[opticajnl]
    \title{Reaching the intrinsic performance limits of superconducting nanowire single-photon detectors up to 0.1~mm wide}

\author{Kristen M. Parzuchowski,\authormark{1,2,$\dag$,*} Eli Mueller,\authormark{1,3,$\dag$} Bakhrom G. Oripov,\authormark{1,2} Benedikt Hampel,\authormark{1,2} Ravin A. Chowdhury,\authormark{1,2} Sahil R. Patel,\authormark{4,5} Daniel Kuznesof,\authormark{6} Emma K. Batson,\authormark{1,7} Ryan Morgenstern,\authormark{1,2} Robert H. Hadfield,\authormark{6} Varun B. Verma,\authormark{1} Matthew D. Shaw,\authormark{5} Jason P. Allmaras,\authormark{5} Martin J. Stevens,\authormark{1} Alex Gurevich,\authormark{8} Adam N. McCaughan\authormark{1,2}}

\address{
\authormark{1}Physical Measurement Laboratory, National Institute of Standards and Technology, 325 Broadway, Boulder, Colorado 80305, USA\\

\authormark{2}Department of Physics, University of Colorado, 2000 Colorado Avenue, Boulder, Colorado 80309, USA\\

\authormark{3}Department of Electrical Engineering, University of Colorado, 1200 Larimer Street, Denver, Colorado 80204, USA\\

\authormark{4}Department of Applied Physics and Materials Science, California Institute of Technology, 1200 E. California Boulevard, Pasadena, California 91125, USA\\

\authormark{5}Jet Propulsion Laboratory, California Institute of Technology, 4800 Oak Grove Drive, Pasadena, California 91109, USA\\

\authormark{6}James Watt School of Engineering, University of Glasgow, University Avenue, Glasgow G12 8QQ, Scotland, UK\\

\authormark{7}Department of Electrical Engineering and Computer Science, Massachusetts Institute of Technology, 77 Massachusetts Avenue, Cambridge, Massachusetts 02139, USA\\

\authormark{8}Department of Physics, Old Dominion University, 5155 Hampton Boulevard, Norfolk, Virginia 23529, USA\\

\authormark{$\dag$}These authors contributed equally to this work.}

\email{\authormark{*}kristen.parzuchowski@nist.gov}

\begin{abstract*}
Superconducting nanowire single-photon detectors combine high detection efficiency, low noise, and excellent timing resolution, making them a leading platform for photon-counting applications. However, despite decades of materials and fabrication research, detector performance has never been shown to match theoretical performance expectations. Here, we demonstrate for the first time in situ tuning of a detector from its typical, suboptimal operation, to a regime limited only by material quality, allowing the device to reach its intrinsic performance limit. Our approach is based on current-biased superconducting “rails” placed on either side of the detector that redistribute current across its width to achieve peak performance. This technique reduces the dark count rate by ten orders of magnitude. Further, we show operation at this intrinsic performance limit for devices up to 0.1~mm wide, and also demonstrate near-unity internal detection efficiency at a wavelength of 4~$\mu$m for a 20~$\mu$m-wide detector---a factor of 20 wider than the current state of the art. This work enables future detectors to overcome the Pearl limit for device width, paving the way for arbitrarily large detectors. 
\end{abstract*}

\section{Introduction}\label{sec1}
Single-photon detection underpins a wide range of emerging photonic technologies, from quantum information processing~\cite{white2025robust,dalbec2025accurate} and secure communications~\cite{grunenfelder2023fast,jabir2025precision} to photon-starved biomedical imaging~\cite{wang2024comprehensive,hughes2026superconducting}. Among the available single-photon detector technologies, superconducting nanowire single-photon detectors (SNSPDs) are highly sought after for these applications, offering detection efficiencies exceeding 98\%~\cite{reddy2020superconducting,hu2020detecting,chang2021detecting}, dark count rates under 1~cnt/day~\cite{chiles2022new}, timing jitter below 3~ps~\cite{korzh2020demonstration}, wire widths up to 60~$\mu$m~\cite{zhang2021physical,yabuno2023superconducting,yabuno2025two}, and sensitivity extending into the mid-infrared~\cite{taylor2023low,hampel2026tungsten,wang2026saturated}. While these metrics are remarkable, they have each been optimized in isolation, with device designs tailored to enhance individual performance metrics at the expense of others. Numerous efforts to improve all-around device performance have focused on identifying promising superconducting films and then refining the fabrication protocols specific to each film. This time-consuming process has yielded real progress, however it is generally incremental relative to theoretical performance expectations~\cite{vodolazov2017single}. A central, but often overlooked reason for this performance gap is the limited operating current of the detector.

Among nominally identical SNSPDs, the highest performing devices are those that can be operated at the highest bias current. Although the theoretical upper bound on the total supercurrent is set by the depairing current $I_{\mathrm{d}}$~\cite{tinkham2004introduction}, real devices transition to the resistive state at a reduced current known as the switching current $I_{\mathrm{sw}}$. Even in state-of-the-art SNSPDs, the ratio $I_{\mathrm{sw}}/I_{\mathrm{d}}$, sometimes called the constriction factor, rarely exceeds 0.7~\cite{frasca2019determining,chiles2020superconducting,haneishi2024evaluation}. 

The primary source of low $I_{\mathrm{sw}}/I_{\mathrm{d}}$ ratios is generally considered to be current pileup at the device edges. Current crowding lowers the energy barrier for vortex entry at the device boundaries, and consequently increases the dark count rate. While current crowding in SNSPDs is often associated with sharp bends or corners in meander structures~\cite{clem2011geometry}, a simple straight superconducting wire also exhibits a nonuniform current distribution across its width, with current density lowest in the middle and peaked near the edges. The mechanism of this edge current crowding is two-fold: First, real devices inevitably have lithographic defects along the edge of the wire, which lead to geometrical current crowding. Second, the Meissner effect causes a non-uniform flow of sheet current density $J(x)$ across the wire width, with $J(x)$ being lowest at the center and maximum at the edges. This current non-uniformity increases with the width of the wire $w$~\cite{haneishi2024evaluation} and is generally seen as the fundamental barrier that prevents the development of SNSPDs with wire widths approaching the scale of the magnetic Pearl length $\Lambda$~\cite{pearl1964current}, typically on the order of 100's of $\mu$m in thin-film SNSPD materials~\cite{haneishi2024evaluation,zhang2016characteristics}. Numerous strategies aimed at reducing the contribution of current crowding have been implemented~\cite{baghdadi2021enhancing,yabuno2023superconducting,strohauer2025current,yabuno2025two}, yielding significant performance gains over the years. However, never has there been a clear indicator that the detectors have reached their intrinsic limit, characterized by the onset of dark counts due to unbinding of vortex-antivortex (VAV) pairs~\cite{halperin1979resistive}. For a given material, this limit is fundamental---it depends only on temperature and superconducting properties of the superconducting film.

Operating at currents far below $I_{\mathrm{d}}$ can be detrimental to photon detection, especially at low photon energies. As the operating current is reduced, higher photon energies are necessary to reach the threshold to form a critical hotspot across the wire width. Because this threshold energy scales with the cross sectional area of the detector, wide SNSPDs are typically inefficient for infrared detection. However, theoretical models predict that as $I_{\mathrm{sw}}$ approaches $I_{\mathrm{d}}$, the minimum detectable photon energy becomes only weakly dependent on wire width~\cite{vodolazov2017single}, allowing the possibility to develop SNSPDs with high efficiency into the infrared and ultra-wide widths surpassing the Pearl length. Here we refer to these detectors as wide SNSPDs, retaining the widely recognized acronym. Although these detectors employ micrometer-scale rather than nanometer-scale widths, the acronym is used to ensure broad understanding of the detector type and its functionality. Scaling up the size of SNSPDs simplifies fabrication, enables higher signal amplitudes, unlocks polarization-insensitivity, and facilitates efficient free-space coupling.

\begin{figure}[htbp]
\centering
\includegraphics[width=0.97\textwidth]{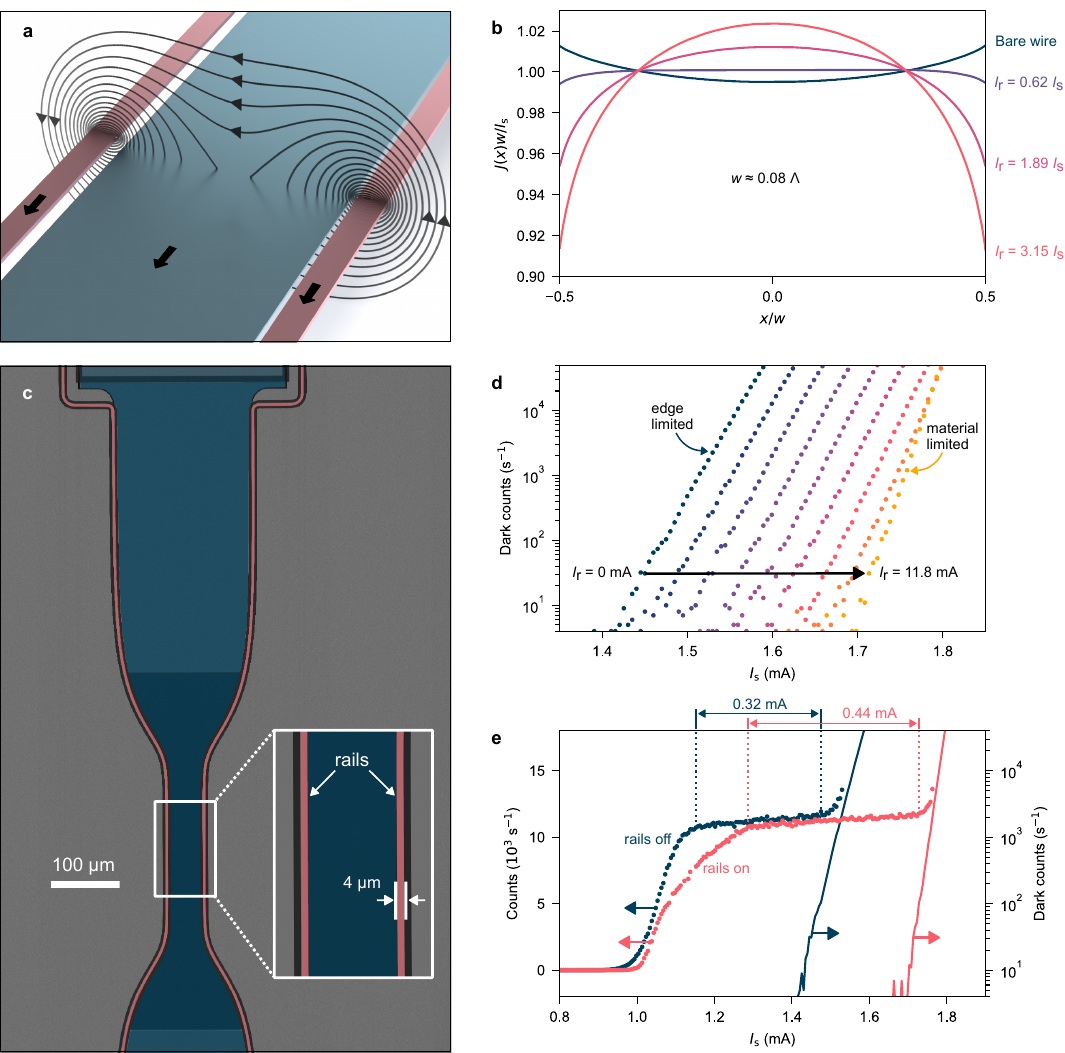}
\caption{(a) Illustration of an SNSPD (blue) integrated with adjacent rails (pink). Simulated magnetic field lines illustrate how the rails modify the self-induced magnetic field of the SNSPD. (b) Normalized current density as a function of position $x$ along the wire width $w$ simulated for a bare wire without rails (dark blue) and for a wire tuned by a series of increasing rail currents (purple to pink). For these simulations, the wire width is $w = 0.08 \Lambda$ and the rails of width $0.15w$ are displaced from the edges of the wire horizontally and vertically by $w/200$. (c) False-color SEM image of 50~$\mu$m-wide WSi wire (dark blue) with 4~$\mu$m-wide Nb rails (pink). The WSi in the wide regions of the taper (lighter blue) is covered with an isolated layer of Nb. The inset shows a zoomed-in image of the constriction. (d) Measured dark count rate as a function of $I_{\mathrm{s}}$ for a 50~$\mu$m-wide wire for a series of increasing $I_{\mathrm{r}}$ values from 0~mA (dark blue) to 11.8~mA (yellow). Dark counts onset at higher $I_{\mathrm{s}}$ as $I_{\mathrm{r}}$ is increased. The blackbody background in this measurement was below approximately 4~s$^{-1}$, which corresponds to the lower limit of the y-axis and is therefore not visible in the plot (e) Measured flood-illuminated 1550~nm photon count rate (left vertical axis, data points) and dark count rate (right vertical log-scale axis, lines) as a function of $I_{\mathrm{s}}$ for rails off (dark blue) and on (pink) for a 50~$\mu$m-wide wire. Gold covers were placed over the tapers of the SNSPD to prevent detection events outside the constriction. The plateau width is $\approx 320\, \mu$A with rails turned off and extends to $\approx 440\, \mu$A with rails turned on. The start of the plateau was set as the value of $I_{\mathrm{s}}$ at which the count rate exceeds $10.7\times10^3$~s$^{-1}$. The end of the plateau was set as the value at which the dark count rate exceeds 100~s$^{-1}$.}\label{fig1}
\end{figure}

In this work, we introduce a method to in situ tune $I_{\mathrm{sw}}$ closer to $I_{\mathrm{d}}$, thereby enabling the device to operate at its intrinsic performance limit. Our method relies on placing an SNSPD between two current-carrying superconducting ``rails", as shown in Fig.~\ref{fig1}(a). Here, the magnetic self-field of the rails partly cancels the perpendicular component of the self-field of the SNSPD, reducing the current density $J(x)$ at the edges~\cite{Newhouse1969,cadorim2024harnessing}. This reduction inverts the typical $J(x)$ profile of a superconducting wire, yielding minima at the edges and a maximum at the center, thereby increasing the energy barrier for vortex entry from the edge of the wire. Figure~\ref{fig1}(b) shows the evolution of $J(x)$ calculated by solving the London equation for different rail currents $I_{\mathrm{r}}$ (see supplementary information). The SNSPD-rail architecture not only eliminates the edge current crowding relative to that of a bare SNSPD without rails, but also allows in situ tuning of $J(x)$ at the edge by varying $I_{\mathrm{r}}$ to reach the max
imum performance. For instance, Fig.~S2 in the supplementary information shows tunable $J(x)$ profiles calculated for a 3~nm thick and 2.5~mm-wide WSi wire that is 4 times larger than the Pearl length.

To measure the efficacy of this technique, we measured a variety of SNSPD-rail devices composed of thin-film WSi ($\Lambda = 600\pm30\,\mu$m) SNSPDs with adjacent Nb rails. By turning on the rails, we demonstrate a 100~$\mu$m-wide SNSPD with an over 10 orders-of-magnitude reduction in dark counts and a detection plateau at 1550~nm widened by more than $40 \%$. In 20~$\mu$m-wide SNSPDs, we reached near-unity internal detection efficiency (IDE) of 4~$\mu$m photons and lowered detector jitter by $\approx 30 \%$. Furthermore, the rails allowed us to recover a detection plateau at 1550~nm in a device that would otherwise not be photosensitive because of its low $I_{\mathrm{sw}}/I_{\mathrm{d}}$ ratio. These significant improvements were made without compromising the photon detection efficiency across the width of the SNSPD. Our results show that the rails tune the device performance from an edge-limited to a material-limited dark count state. To our knowledge, this is the first demonstration of in situ tuning of an SNSPD to its intrinsic performance limit and establishes a pathway to scaling high-performing SNSPDs to widths exceeding the Pearl length. 

\section{Results}\label{sec2}
Our SNSPD-rail architecture is illustrated in Figure~\ref{fig1}(a). An SEM image of a fabricated 50~$\mu$m-wide WSi SNSPD with 4~$\mu$m-wide adjacent Nb rails is shown in Figure~\ref{fig1}(c). Here the $\approx 50$~nm-thick rails are displaced $\approx 150$~nm from the edge of the SNSPD. The $\approx 3$~nm-thick WSi wire has a constriction length of 100~$\mu$m. We fabricated (see details in Methods) and tested numerous devices varying in SNSPD width from 1~$\mu$m to 100~$\mu$m. Device constriction length varied from 10~$\mu$m to 300~$\mu$m, which is mentioned only for clarity as length does not play a significant role in the results or analysis that follows. Here we present representative experimental results from these devices. Unless otherwise specified, all measurements were performed in a fiber-coupled cryostat at $T=$~900~mK. The device chip was enclosed in an aluminum box at the low-temperature stage to minimize blackbody background. The output of the optical fiber (SMF-28) was coupled through a fiber port mounted on the aluminum box, and was offset from the device chip by about 5 cm, thus producing flood illumination over the entire 1~cm$^2$ chip. For detector pulse readout, we used a wide-band bias tee (Mini-Circuits ZFBT-4R2GW-FT+) and two low noise amplifiers (Mini-Circuits ZFL-1000LN+ and ZX60-P103LN+).

\subsection{Suppression of dark counts}
Figure~\ref{fig1}(d) shows the dark count rate as a function of the SNSPD bias current $I_{\mathrm{s}}$ for a series of $I_{\mathrm{r}}$ values for a 50~$\mu$m-wide wire. The dark count rate follows a log-linear dependence on $I_{\mathrm{s}}$, consistent with dark counts initiated by Arrhenius thermally-activated vortex crossings~\cite{bulaevskii2011vortex} as described in the supplementary information. Increasing $I_{\mathrm{r}}$ from 0~mA up to 11.8~mA shifts the dark count rate curve to higher $I_{\mathrm{s}}$ values. For $I_{\mathrm{r}}>11.8$~mA (not shown), this trend reverses and the curve shifts to lower $I_{\mathrm{s}}$. Here, $I_{\mathrm{r}}^{*}\approx11.8$~mA represents the optimal rail current that maximizes $I_{\mathrm{sw}}$ of the SNSPD, where $I_{\mathrm{sw}}$ is defined operationally as the value of $I_{\mathrm{s}}$ at which the dark count rate exceeds 100~s$^{-1}$. The integration time for this measurement was 1~s, which limits the minimum measurable dark count rate to 1~s$^{-1}$. We extrapolated the $I_{\mathrm{r}}^{*}$ dark count curve to the $I_{\mathrm{s}}$ value at which the dark counts are 1000~s$^{-1}$ for $I_{\mathrm{r}}=0$ and estimate a more than 9 orders of magnitude reduction in dark counts at $I_{\mathrm{r}}^{*}$ relative to $I_{\mathrm{r}}=0$. At $I_{\mathrm{r}} \approx I_{\mathrm{r}}^{*}$, there is an abrupt increase in the slope of the dark count rate curve that persists for $I_{\mathrm{r}}>I_{\mathrm{r}}^{*}$. As will be discussed below, this slope increase is indicative of a transition from a regime where dark counts are dominated by penetration of vortices from the edges ($I_{\mathrm{r}}<I_{\mathrm{r}}^{*}$), to a regime where dark counts are dominated by unbinding of VAV pairs in the bulk material ($I_{\mathrm{r}}>I_{\mathrm{r}}^{*}$). Here, the intrinsic performance limit has been reached when $I_{\mathrm{r}}=I_{\mathrm{r}}^*$, and further increases in $I_{\mathrm{r}}$ result in a gradual loss of peak performance due to an increasingly nonuniform current density. Note that reversing the rail current polarity such that $I_{\mathrm{s}}$ and $I_{\mathrm{r}}$ flow in opposite directions shifts the dark count rate curve to lower $I_{\mathrm{s}}$ values, thereby degrading device performance. 

\subsection{Impact on photosensitivity}
Figure~\ref{fig1}(e) shows the count rate as a function of $I_{\mathrm{s}}$ for the same 50~$\mu$m-wide device under flood illumination of $\lambda=1550$~nm light with the rails off ($I_{\mathrm{r}} = 0$~mA) and with the rails on ($I_{\mathrm{r}} = I_{\mathrm{r}}^{*}$). The count rate reaches a detection plateau, indicating saturated IDE, both with the rails on and off. Note that with the rails on, the onset of photosensitivity shifts to higher $I_{\mathrm{s}}$ values compared to that with the rails off; however, the dark-count onset shifts by an even larger amount, resulting in a $\approx 40\%$ increase in the detection plateau length. We also observed that the photosensitivity onset with the rails on deviates from a typical sigmoid shape. We review both of these photosensitivity-onset features in the discussion section.

\begin{figure}[bt]
\centering
\includegraphics[width=0.99\textwidth]{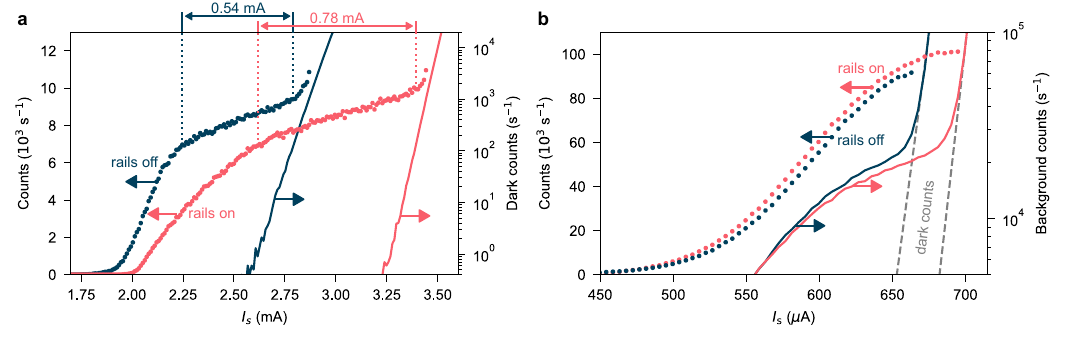}
\caption{(a) Measured flood-illuminated 1550~nm photon count rate (left vertical axis, data points) or dark count rate (right vertical log-scale axis, lines) as a function of $I_{\mathrm{s}}$ for rails off (dark blue) or rails on (pink) for a 100~$\mu$m-wide wire. The plateau width is $\approx 540\, \mu$A with rails turned off and extends to $\approx 780\, \mu$A with rails turned on. The start of the plateau was set as the value of $I_{\mathrm{s}}$ at which the count rate exceeds 6.9$\times10^3$~s$^{-1}$. The end of the plateau was set as the value at which the dark count rate exceeds 100~s$^{-1}$. (b) Measured background-subtracted, flood-illuminated 4~$\mu$m photon count rate (left vertical axis, data points) and background count rate (right vertical log-scale axis, lines) for a 20~$\mu$m-wide wire as a function of $I_{\mathrm{s}}$ for rails off (dark blue) and rails on (pink). At low $I_{\mathrm{s}}$ the background counts are dominated by blackbody background photons entering through the optical window on the cryostat, then, as $I_{\mathrm{s}}$ is increased, the exponential dark counts takeover as indicated by the dashed gray line. Turning on the rails shifts the dark counts to higher $I_{\mathrm{s}}$ values enabling measurements at higher $I_{\mathrm{s}}$ values. In this region, a rollover of counts is observed, indicating the start of a plateau and near-unity IDE.}\label{fig2}
\end{figure}

Similarly, Fig~\ref{fig2}(a) shows the count rate as a function of $I_{\mathrm{s}}$ for a 100~$\mu$m-wide device with 10~$\mu$m constriction length under flood illumination of $\lambda=1550$~nm light with the rails off ($I_{\mathrm{r}} = 0$~mA) and with the rails on ($I_{\mathrm{r}} = I_{\mathrm{r}}^{*} \approx 11.1$~mA). For this measurement, the pulse amplitude was so large that amplifiers were not used. The count rate exhibits a slope on the plateau that we attribute to photon detection events in the taper of the device. In contrast, the plateau in Fig.~\ref{fig1}(e) is flat, as the tapers in that device were blocked with a gold layer. Similar to Fig.~\ref{fig1}(e), turning the rails on shifts the onset of photosensitivity to higher $I_{\mathrm{s}}$ values, while the dark-count onset shifts by a larger amount, resulting in a $\approx 44\%$ increase in the detection plateau length. We extrapolated the $I_{\mathrm{r}}^{*}$ dark count curve to the $I_{\mathrm{s}}$ value at which the dark counts are 1000~s$^{-1}$ for $I_{\mathrm{r}}=0$ and estimate a more than 10 orders of magnitude reduction in dark counts at $I_{\mathrm{r}}^{*}$ relative to $I_{\mathrm{r}}=0$.

To test the upper bound on the wavelength sensitivity of our SNSPD-rail devices, we tested a 20~$\mu$m-wide SNSPD with 300~$\mu$m constriction length in a $\approx 260$~mK cryostat~\cite{hampel2026tungsten} fit with an optical window for free-space coupled infrared light sources, which in this case was a blackbody source with a temperature of 630~K, and a monochromator at the 40~K stage used for precise wavelength selection. This optical window resulted in higher background photon counts than the fiber-coupled experiments at 1550~nm [Fig.~\ref{fig1}(d) and Fig.~\ref{fig2}(a)]. Figure~\ref{fig2}(b) shows the background-subtracted photon count rate of the 20~$\mu$m-wide SNSPD under flood illumination of $\lambda= 4\,\mu$m photons as a function of $I_{\mathrm{s}}$ with the rails off ($I_{\mathrm{r}} = 0$~mA) and with the rails on ($I_{\mathrm{r}} = I_{\mathrm{r}}^{*} \approx 10.1$~mA). The background counts due to this light source and cryostat are evident in the $I_{\mathrm{s}}=550\,\mu$A to $675\,\mu$A range before the log-linear dark counts dominate. By turning on the rails, the dark counts are reduced enabling measurement of the 20~$\mu$m-wide wire's photon count rate at higher $I_{\mathrm{s}}$. The count rate at higher $I_{\mathrm{s}}$ increases and rolls over onto the start of a detection plateau, indicating near-unity IDE. This result represents a factor of 20 increase in wire width compared to the current state of the art for high efficiency SNSPDs at mid-infrared wavelengths~\cite{wang2026saturated}. We note that during this measurement, biasing the rails at $I_{\mathrm{r}}^{*}$ increased the operating temperature by approximately 120~mK due to Joule heating, likely due to the high-resistivity CuNi coaxial cabling at the low-temperature stage. This heating could be mitigated by instead using superconducting coaxial cables.

\begin{figure}[bt]
\centering
\includegraphics[width=0.49\textwidth]{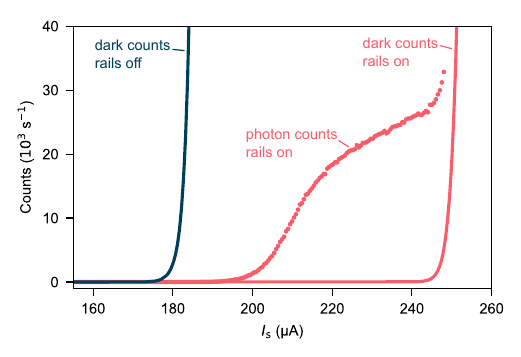}
\caption{Measured flood-illuminated 1550~nm photon count rate (pink data points) and dark count rate (lines) as a function of $I_{\mathrm{s}}$ for rails off (dark blue) and rails on (pink) for a 10~$\mu$m-wide wire with low $I_{\mathrm{sw}}/I_{\mathrm{d}}$ ratio. With the rails off, the dark counts begin to dominate at such a low $I_{\mathrm{s}}$ that photon counts cannot be measured. With the rails on, a huge shift of dark counts to higher $I_{\mathrm{s}}$ is observed, and photon counts are measured with near-unity IDE. The slope on the plateau is consistent with photon detection events in the taper of the device.}\label{fig3}
\end{figure}

To determine if the rails could recover an otherwise non-functional detector, we also tested several devices that showed unusually low $I_{\mathrm{sw}}/I_{\mathrm{d}}$ ratios. We attribute the poor performance of these outlier devices to severe lithographic defects~\cite{silhanek2025}, which are inevitably present in a small fraction of devices in any lithographic process. In Fig.~\ref{fig3}, the measured flood-illuminated 1550~nm photon count rate and dark count rate is plotted as a function of $I_{\mathrm{s}}$ for one of these devices. This device is 10~$\mu$m wide with a 300~$\mu$m constriction length. With the rails off ($I_{\mathrm{r}}=0$~mA), $I_{\mathrm{sw}}/I_{\mathrm{d}}\approx44$\% and the dark counts dominate at such a low $I_{\mathrm{s}}$ values that the wire has no photosensitivity to 1550~nm photons. With the rails on ($I_{\mathrm{r}}=20.9$~mA), the dark count curve shifts dramatically to higher $I_{\mathrm{s}}$ values, reaching $I_{\mathrm{sw}}/I_{\mathrm{d}}\approx63$\%, and photon counts can be measured with near-unity IDE. Thus, the rails recovered the performance of the SNSPD that otherwise would have been non-functional at 1550~nm. 

\subsection{Impact on timing jitter}
\begin{figure}[tb]
\centering
\includegraphics[width=0.99\textwidth]{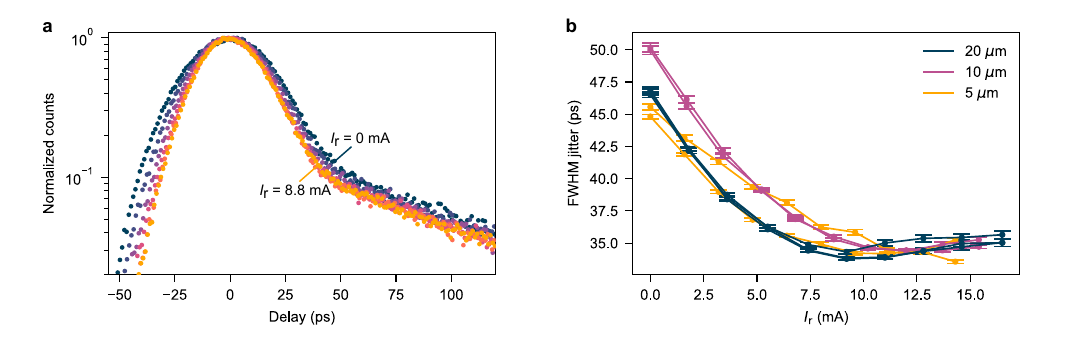}
\caption{(a) Jitter histograms for a 20~$\mu$m-wide wire under 1550~nm flood illumination for six $I_{\mathrm{r}}$ values ranging from $I_{\mathrm{r}}=0$ to $I_{\mathrm{r}}=8.8$~mA. The long tail onsetting at $\approx 40$~ps delay time is likely due to detection events in the taper. The jitter histogram narrows as $I_{\mathrm{r}}$ is increased. (b) FWHM system jitter as a function of $I_{\mathrm{r}}$ for two 5~$\mu$m (yellow), two 10~$\mu$m (pink) and three 20~$\mu$m (dark blue) wide wires. As $I_{\mathrm{r}}$ is increased, the jitter decreases until it reaches a minimum.}\label{fig4}
\end{figure}
Figure~\ref{fig4}(a) shows the histogram of detection latency for a 20~$\mu$m-wide wire with 300~$\mu$m constriction length under 1550~nm flood illumination for six $I_{\mathrm{r}}$ values ranging from $I_{\mathrm{r}}=0$ to $I_{\mathrm{r}}=8.8$~mA. The long tail is likely due to detection events in the taper of our wires as has been noted in previous studies~\cite{korzh2020demonstration}. As $I_{\mathrm{r}}$ is increased, the histogram narrows. We suspect that this may be caused by the merging of two latency distributions: one originating from edge vortex-mediated detections and another from VAV-pair-mediated detections, however a more thorough study is necessary to confirm this hypothesis.

In Fig.~\ref{fig4}(b) the full width at half maximum (FWHM) system jitter is plotted as a function of $I_{\mathrm{r}}$ for several devices with 300~$\mu$m constriction lengths and widths of 5~$\mu$m, 10~$\mu$m and 20~$\mu$m. There is no clear dependence of jitter on wire width; however, all wires demonstrate improved jitter upon application of rail current until an optimum value is reached. This optimum jitter value typically occurred at an $I_{\mathrm{r}}$ value that differed from $I_{\mathrm{r}}^*$ by a few milliamperes. After this optimum jitter value, the jitter slightly increases as the rail current is increased further. Interestingly, the optimum jitter of all three wire widths is similar, $\approx 35$~ps. 

We measured several contributions to the system jitter, including the time tagger single-channel jitter, the laser sync jitter, the sync-to-optical jitter and the laser period jitter, which have FWHM contributions of 13.6~ps, 5.6~ps, 2.1~ps and $\approx 3.4$~ps respectively. The remaining contributions are from the noise jitter of the SNSPD pulse~\cite{caloz2019intrinsically}, the intrinsic variations in the generation of a hotspot and the geometric jitter of the SNSPD. We estimate the standard deviation of the longitudinal geometric jitter to be 10.0~ps, 7.9~ps and 6.3~ps for the 5~$\mu$m, 10~$\mu$m and 20~$\mu$m-wide devices, respectively. Further study is required to fully quantify the total geometric jitter.

\subsection{Tunable proximity to the depairing current}
\begin{figure}[tb]
\centering
\includegraphics[width=0.49\textwidth]{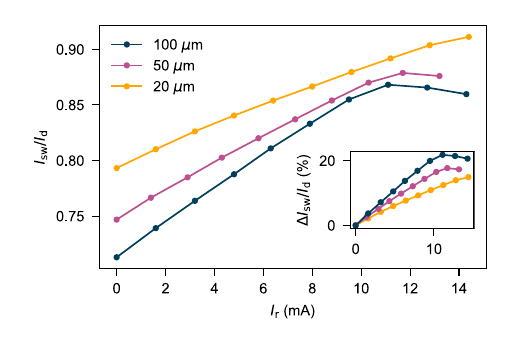}
\caption{Measured $I_{\mathrm{sw}}/I_{\mathrm{d}}$ as a function of $I_{\mathrm{r}}$ for wire widths of 20~$\mu$m (yellow), 50~$\mu$m (pink) and 100~$\mu$m (dark blue). As $I_{\mathrm{r}}$ is increased, $I_{\mathrm{sw}}/I_{\mathrm{d}}$ also increases until it reaches $I_{\mathrm{r}}^*$. Percent change ($\Delta$) in $I_{\mathrm{sw}}/I_{\mathrm{d}}$ as a function of $I_{\mathrm{r}}$ is shown in the inset for the same devices. As wire width is increased, larger increases in $I_{\mathrm{sw}}/I_{\mathrm{d}}$ are measured.}\label{fig5}
\end{figure}
As a demonstration of how the rails impact the switching current, Fig.~\ref{fig5} shows $I_{\mathrm{sw}}/I_{\mathrm{d}}$ as a function of $I_{\mathrm{r}}$ for a 20~$\mu$m-wide device with 100~$\mu$m constriction length, a 50~$\mu$m-wide device with 100~$\mu$m constriction length, and a 100~$\mu$m-wide device with 10~$\mu$m constriction length. Here $I_{\mathrm{sw}}$ is defined as the value of $I_{\mathrm{s}}$ at which the dark count rate exceeds 100~s$^{-1}$. For all wire widths, $I_{\mathrm{sw}}/I_{\mathrm{d}}$ increases for positive $I_{\mathrm{r}}$ and reaches a peak over the range of $I_{\mathrm{r}}$ measured. This peak occurs at $I_{\mathrm{r}}^*$. Although the 20~$\mu$m wire data was not measured past the peak, it was confirmed that the maximum $I_{\mathrm{r}}$ value tested corresponded to the point at which the slope of the dark count rate versus $I_{\mathrm{s}}$ curve steepened, suggesting the device has reached its performance limit. As wire width increases, the change in $I_{\mathrm{sw}}/I_{\mathrm{d}}$ with $I_{\mathrm{r}}$ becomes more significant. This trend is consistent with our calculations that show that wider wires with correspondingly stronger edge current crowding benefit more from its suppression by the rails.

To determine the depairing current of our film, which was used in Fig.~\ref{fig5}, we measured the change in kinetic inductance of step-impedance resonators as a function of bias current as described in Ref.~\cite{frasca2019determining}. These resonators were fabricated from a nominally identical film to that used for the SNSPD-rail devices. The measured kinetic inductance change was fit to the fast and slow relaxation models described in Ref.~\cite{clem2012kinetic}. We found that the fast relaxation model was a better fit for all resonators. These fits were used to derive the depairing current per unit of wire width with an average of 39~$\mu$A~$\mu$m$^{-1}$. 

\section{Discussion}
Our results show that SNSPD-rail devices reverse current crowding at the wire edges and remove the fundamental Pearl length limitation on wire width. This also follows from our numerical calculations that show that the current density at the edges can be reduced lower than that in the center of the wire even for a SNSPD with $w \gg \Lambda$. Moreover, the required reduction of $J(x)$ at the edges is achieved with a normalized rail current $I_\mathrm{r}/I_\mathrm{s}$ that diminishes as the wire width increases from $w \ll \Lambda$ to $w > \Lambda$. Although some superconducting devices suffer consequences from the linear increase in magnetic flux trapping with device area, here, the direct current flowing through the SNSPD flushes out the flux trapped during the cooldown, yielding no consequences. We anticipate that the next limitation to scale to wider SNSPD widths may be set by a degradation in film quality over large length scales. 

We observed that for all devices, $I_{\mathrm{sw}}(w)$ are maximized at optimum rail currents $I_{\mathrm{r}}^*(w)$. The slopes of the log-linear dark count rates as a function of $I_{\mathrm{s}}$ (such as that shown in Fig.~\ref{fig1}(d)) increase abruptly at $I_{\mathrm{r}} = I_{\mathrm{r}}^*$. The dark count rate $S$ generally follows a thermally-activated Arrhenius behavior caused by penetration of vortices from the edge given by $S_{\mathrm{e}}(I_{\mathrm{s}}) \simeq S_{\mathrm{e0}}\times\exp\!\left[-(1 - I_{\mathrm{s}}/I_{\mathrm{d}})U_{\mathrm{e}}/T\right]$ where $U_{\mathrm{e}}$ is the energy barrier for vortex entry. For the data $I_{\mathrm{r}} < I_{\mathrm{r}}^*$, we have $U_{\mathrm{e}} \approx 83\, \mathrm{K}$ for the $50\,\mu\mathrm{m}$-wide wire shown in Fig.~\ref{fig1}(d) and $U_{\mathrm{e}} \approx 71\, \mathrm{K}$ for the $100\,\mu\mathrm{m}$-wide wire shown in Fig.~\ref{fig2}(a). These values of $U_{\mathrm{e}}$ are consistent with our evaluation of the vortex energy barrier for penetration of vortices from the edges given in the supplementary information. In turn, the nearly doubling of the logarithmic slope observed at $I_{\mathrm{r}} > I_{\mathrm{r}}^*$ is strong experimental evidence that the dark count rate becomes dominated by the unbinding of VAV pairs of radius $\sim \xi J_{\mathrm{d}} / J \ll w$ in the center of the wire. This is because the energy of a vortex spaced by $u$ from an ideal edge of a wire is half the energy of a VAV pair of radius $u$ in the bulk of a wire, so that now the dark count rate is given by $S_{\mathrm{b}}(I) \simeq S_{\mathrm{b0}}\exp\!\left[-(1 - I_{\mathrm{s}}/I_{\mathrm{d}})U_{\mathrm{b}}/T\right]$ where $U_{\mathrm{b}} = 2U_{\mathrm{e}}$. 

As shown in the supplementary information, the ratio of the logarithmic slopes $\partial_{I_\mathrm{s}}\mathrm{ln}S$ for vortex hopping from the edges and for VAV pair production are related by $\partial_{I_\mathrm{s}}\mathrm{ln}S_{\mathrm{b}}/\partial_{I_\mathrm{s}}\mathrm{ln}S_{\mathrm{e}}\simeq 2 \tilde{J}_{\mathrm{d}}/J_{\mathrm{d}}$ where $J_{\mathrm{d}}$ and $\tilde{J}_{\mathrm{d}}$ are the mean depairing current densities in the bulk and at the edges, respectively. Here $\tilde{J}_{\mathrm{d}}$ is reduced relative to $J_{\mathrm{d}}$ by the current crowding at lithographic defects. For instance, the slope ratio $\approx 1.79$ observed in the $50\,\mu\mathrm{m}$-wide wire shown in Fig.~\ref{fig1}(d) implies $\tilde{J}_{\mathrm{d}} = 0.895 J_{\mathrm{d}}$ and the ratio $\approx 1.53$ observed in the $100\,\mu\mathrm{m}$-wide wire shown in Fig.~\ref{fig2}(a) implies $\tilde{J}_{\mathrm{d}} = 0.765 J_{\mathrm{d}}$. To mitigate penetration of vortices from the edges for these wires, the rails should carry current sufficient to reduce $J(x)$ at the edges by $10.5\%$ and $23.5\%$ respectively.

Two important conclusions follow from our analysis of dark counts: 
(1) In addition to increasing $I_{\mathrm{sw}}$, the rails greatly (up to 10 orders of magnitude) reduce dark counts caused by penetration of vortices from the edges. (2) At $I_{\mathrm{r}}= I_{\mathrm{r}}^{*}$ the wire achieves its maximum $I_{\mathrm{sw}}/I_{\mathrm{d}}$. As stated above, we define $I_{\mathrm{sw}}$ as the current at which the dark count rate equals $100\,\mathrm{\mathrm{s}}^{-1}$. The VAV pair unbinding makes it impossible to reach $I_\mathrm{s}=I_\mathrm{d}$ in a wire without switching to the resistive state, but, the rails do allow the device to operate at its intrinsic performance limit with maximum photon sensitivity. In addition to the fundamental limits of $I_{\mathrm{sw}}$ and the SNSPD sensitivity, which are both caused by the VAV pair production evaluated in the supplementary information, there are also technological and materials science limitations to its maximum value. These limitations result from variations in film thickness, lithographic defects, inhomogeneities of superconducting properties due to local non-stoichiometry, grain boundaries in polycrystalline films, film adhesion with the substrate, etc. Beyond these limitations, our estimated $I_{\mathrm{d}}$ has uncertainty associated with it. For example, our method for estimating $I_{\mathrm{d}}$ of our devices uses assumptions from the BCS model that do not account for the effects of strong electron--phonon pairing of the Eliashberg theory~\cite{nicol1991temperature}, or the reduction of $J_{\mathrm{d}}$ by subgap quasiparticle states~\cite{kubo2020superfluid}, both of which are especially pronounced in thin, superconducting films. Nonetheless, tuning $J(x)$ by rails enables reaching the intrinsic limit of $I_{\mathrm{sw}}$ for a particular SNSPD even if both the exact value of $J_{\mathrm{d}}$ and the extent of reduction of $J_{\mathrm{d}}$ at the edges is unknown.

We observed that current redistribution in SNSPDs was possible without affecting detection efficiency uniformity; the photon count rate on the detection plateau varies by less than 1\% across the full range of $I_{\mathrm{r}}$ as described in the Methods section. However, the results shown in Fig.~\ref{fig1}(e) and Fig.~\ref{fig2}(a) clearly show that the onset of photosensitivity shifts slightly to higher $I_{\mathrm{s}}$ values with the rails on. This behavior may reflect a nonuniform threshold current density for photodetection across the wire in which the photon detections near the edge have a lower detection threshold current density than that in the bulk~\cite{zotova2014intrinsic,renema2015position,wang2015local}. In this scenario, when the rails are turned off, the edges will become photosensitive at lower $I_{\mathrm{s}}$ values relative to that for the bulk. However, when the rails are turned on and the current density near the edges is reduced, a higher total $I_{\mathrm{s}}$ is required for the edge to reach its threshold current density, shifting the onset of photosensitivity to higher $I_{\mathrm{s}}$ values. We also observed a change in shape of the photosensitivity-onset curve with the rails turned on for the measurements at $T\approx0.9$~K under $\lambda= 1550$~nm illumination [Fig.~\ref{fig1}(e) and Fig.~\ref{fig2}(a)], which may also be a result of the reduced current density at the edges of the wire. Notably, this effect, which broadens the onset and causes a deviation from a typical sigmoidal curve, was not observed for the measurements at $T\approx260$~mK under $\lambda= 4\,\mu$m illumination [Fig.~\ref{fig2}(b)]. This suggests that the effect is strongly dependent on photon energy and/or thermal fluctuations as has been reported previously~\cite{hong2025impact,wang2019wavelength,yamashita2010temperature}. However, several mechanisms are known to influence the functional form of the IDE as a function of $I_{\mathrm{s}}$~\cite{kozorezov2017fano}, and distinguishing among the possible mechanisms is beyond the scope of this work.

Turning to the timing jitter observed in our devices, the $\approx$~50~ps jitter on our devices up to 20$\,\mu$m wide warrants further discussion. This result seems inconsistent with models in which components of the VAV pair generated by a photon near the edge and in the center of the wire have different transit times~\cite{vodolazov2017single} so that wider wires would have greater timing jitter. This model, along with previous experiments studying jitter dependence on wire width~\cite{korzh2020demonstration} have primarily focused on nanowires $\approx100$~nm wide. It remains unclear how the jitter scales as the widths increase to 1~$\mu$m-100~$\mu$m. For instance, taking the terminal velocity of a vortex $v_{\mathrm{c}} \simeq$~20~km~s$^{-1}$ measured on Pb film bridges~\cite{embon2017imaging} at $J \approx J_{\mathrm{d}}$, one would expect a geometric jitter of  $\sim w/v_c \sim 1\,\mathrm{ns}$ in a $20\,\mu\mathrm{m}$-wide wire, which is 20 times larger than the observed jitter of $\sim 50\,\mathrm{ps}$ shown in Fig.~\ref{fig4}. The value of $v_{\mathrm{c}}$ observed on Pb is qualitatively consistent with the estimate $v_{\mathrm{c}} \simeq J_{\mathrm{d}}/\eta \simeq \rho_{\mathrm{n}} \xi / 2\mu_{\mathrm{0}} \lambda^2$, where $\eta = \phi_{\mathrm{0}}^2 / 2\pi \xi^2 \rho_{\mathrm{n}}$ is the Bardeen--Stephen vortex drag coefficient, $\rho_{\mathrm{n}}$ is a normal-state resistivity at $T_{\mathrm{c}}$, and $J_{\mathrm{d}} \simeq \phi_{\mathrm{0}} / 4\pi \mu_{\mathrm{0}} \lambda^2 \xi$. The same estimate of $v_{\mathrm{c}}$ for our WSi film with $\xi = 10\,\mathrm{nm}$, $\lambda = 950\,\mathrm{nm}$ and $\rho_{\mathrm{n}} = 2.5\,\mu\Omega\,\mathrm{m}$ yields $v_{\mathrm{c}} \sim \rho_{\mathrm{n}} \xi / 2\mu_{\mathrm{0}} \lambda^2 \sim 11$~km~s$^{-1}$, which again predicts a jitter substantially larger than that observed on the $20\,\mu\mathrm{m}$-wide SNSPD. Thus, real devices significantly outperform the estimates of conventional models of vortex-assisted jitter, showing that our current understanding of the dynamics of a resistive domain produced by photon absorption is still incomplete.

It is worth noting that the ultra-wide SNSPDs studied here showed broad detection plateaus for 1550~nm photons even without applying rail current. In addition, a control device without rails, or otherwise any Nb nearby, also demonstrated a similar detection plateau, indicating that the rail structures are not always required for high efficiency photosensitivity to 1550~nm photons. This result is remarkable, and we attribute this to several factors. First, our use of electron-beam lithography, and the relatively short wire lengths in our devices help minimize edge defects~\cite{gaudio2014inhomogeneous}, resulting in relatively high $I_{\mathrm{sw}}/I_{\mathrm{d}}$ compared to those noted in previous works~\cite{chiles2020superconducting,frasca2019determining,haneishi2024evaluation}. Second, our detectors are made from a highly uniform amorphous superconductor~\cite{marsili2013detecting} and the thin-film WSi recipe has a high silicon content ($\approx$~0.48 Si mole fraction) designed to maximize photosensitivity in wide wires~\cite{chiles2020superconducting}. We also note that our wires do not self reset without shunting the device with an inductor and resistor in parallel with the SNSPD at low temperature. We used a PdAu shunt resistor that was nominally 1~$\Omega$, and a Nb spiral inductor that ranged from  0.2~$\mu$H to 1~$\mu$H depending on the experiment. This shunting requirement is likely generic to ultra-wide SNSPDs and may have caused wide wires to be overlooked in the past. 

Other SNSPD designs have similarities to our SNSPD-rail architecture---superconducting nanowire avalanche detectors (SNAP)~\cite{ejrnaes2007cascade,patel2024improvements} have multiple parallel adjacent wires that frequently employ parallel currents. However, the magnitude of the currents in neighboring wires are all equal, which will have a very small effect on the current distribution of the neighboring wires. As noted above, narrow central wires, like those typically used in SNAP, require higher normalized rail current $I_\mathrm{r}/I_\mathrm{s}$ to achieve the same effect as that for wide central wires. Another similar design is the meander-style SNSPD~\cite{gol2003fabrication,reddy2022broadband} that have multiple parallel adjacent wires flowing antiparallel currents. This design also uses equal current magnitudes and narrow wires, likely yielding no noticeable effect. We also note that the broader concept of engineering edge vortex barriers has previously been explored, for example by intentionally roughening one edge of a superconducting strip to produce asymmetric vortex entry and flux rectification under an applied dc magnetic field~\cite{vodolazov2005superconducting}.

To our knowledge, the widest SNSPDs tested to date are 60~$\mu$m-wide detectors made from approximately 2~nm-thick MoSi films~\cite{zhang2021physical}. These devices demonstrated single-photon sensitivity at wavelengths up to 1550~nm, however they did not show saturated IDE. In two recent works by Yabuno et al.~\cite{yabuno2023superconducting,yabuno2025two}, SNSPDs of width 20~$\mu$m made from NbTiN thin films were demonstrated. In this study, the authors used a “high critical current bank (HCCB)” structure, in which the center of the wire material is irradiated with ions to selectively lower $I_{\mathrm{d}}$ relative to the edges. For a fixed current flowing through the wire, this irradiation allows the center to operate closer to $I_{\mathrm{d}}$ than without irradiation. This approach worked extremely well, demonstrating near-unity IDE at $2\,\mu$m for the 20~$\mu$m-wide wire. However, the HCCB does not suppress the edge current crowding caused by Meissner screening, but instead partially compensates for it by reducing $I_{\mathrm{d}}$ in the center relative to the edges. If the wire width continues to approach and exceed the Pearl length, the current crowding will become more pronounced, and eventually exceed the additional vortex energy barrier provided by the HCCB. Therefore, the fundamental limitation on the maximum size of wire width achievable using this approach is still limited by the Pearl length. 

Finally, in addition to boosting the performance limits of any given SNSPD, the rails can also help resolve the Berezinskii--Kosterlitz--Thouless (BKT) transition in thin films~\cite{halperin1979resistive}. As pointed out above, the abrupt change in the slope of the dark count rate at the optimum rail current $I_{\mathrm{r}}^{*}$ shown in Fig.~\ref{fig1}(d) is a clear experimental indication of the suppression of single-vortex penetration from the edges and a transition to BKT unbinding of VAV pairs inside the film. Thus, using rails can be helpful in fundamental investigations of the BKT transition without the masking effects of edge-induced vortex entry.

\section{Conclusion}\label{sec4}
We have demonstrated, for the first time, a method to in situ tune the performance of an SNSPD to its intrinsic limit, paving the way to scale to higher-performing detectors as well as wire widths exceeding the Pearl length. The approach uses current-biased superconducting rails on each side of the wire to reduce the current density at the edges, allowing the device to operate at a higher fraction of the depairing current. We pushed a 100~$\mu$m-wide SNSPD to its intrinsic performance limit, demonstrating an extension of the detection plateau at 1550~nm and over 10 orders-of-magnitude reduction in dark counts. We demonstrated that rails provide numerous other follow-on performance benefits to SNSPD technology including increased upper bound of wavelength sensitivity, recovery of detector functionality in low performing devices, and improved detector jitter. 

The orders-of-magnitude reduction in dark counts has positive consequences for scaling these detectors into arrays. In an array, the total dark count rate is multiplied by the number of individual detectors, and thus can quickly add up to a sizeable number of dark counts~\cite{oripov2023superconducting}. With rails, the dark counts can easily be reduced to a nearly negligible rate. 

The ability to scale to wide wires is also important and previously was inaccessible. Wide wires eliminate the need to meander a detector to create a high fill factor~\cite{reddy2022broadband}. With a focused or fiber-coupled beam, all of the photons of interest can be collected onto our 100~$\mu$m-wide detector. Scaling these wide wires up into arrays, the fill factor can be much higher than previous demonstrations~\cite{oripov2023superconducting}. Furthermore, meandered detectors are sensitive to the polarization of light, with light polarized parallel to the straight sections of the detector being less efficiently detected. In contrast, our wide detectors consist of a single wire that can cover the entire focused beam spot, allowing maximum optical efficiency for all polarizations. 

In future studies, we foresee being able to scale these detectors to widths wider than the Pearl length, which could yield high efficiency free-space coupled SNSPDs---eliminating the need for lossy fiber optics. These detectors could immediately be applied to quantum information experiments where loss is detrimental to the fidelity of the transmitted quantum state~\cite{white2025robust,jabir2025precision}. 



\subsection*{Acknowledgements}
This research was funded by NIST (https://ror.org/05xpvk416), University of Colorado Boulder (https://ror.org/02ttsq026), University of Colorado Denver (https://ror.org/02hh7en24), NASEM (https://ror.org/02eq2w707), and the DARPA DSO Synthetic Quantum Nanostructures program. The U.S. Government is authorized to reproduce and distribute reprints for governmental purposes notwithstanding any copyright annotation thereon. DK thanks the EPSRC and SFI Centre for Doctoral Training in Photonic Integration for Advanced Data Storage\\ (CDT-PIADS EP/S023321/1). Certain commercial equipment, instruments, or materials are identified in this paper in order to specify the experimental procedure adequately. Such identification is not intended to imply recommendation or endorsement by NIST, nor is it intended to imply that the materials or equipment identified are necessarily the best available for the purpose.

\subsection*{Disclosures} 
The authors declare no conflicts of interest.




    \putbib[bib]
  \end{bibunit}
\fi



\ifnum\doSupp=1
  \ifnum\doMain=1
    \clearpage
        \begin{center}
      \vspace*{2cm}
      {\Large\bfseries Supplementary Information\par}
    \end{center}
  \fi

  \setcounter{figure}{0}
  \renewcommand{\thefigure}{S\arabic{figure}}

  \setcounter{equation}{0}
  \renewcommand{\theequation}{S\arabic{equation}}

  \begin{bibunit}[opticajnl]
    \section{Methods}\label{secA1}
\subsection{{Fabrication}}
The chips were fabricated with 5 layers: 1 WSi layer, 1 Nb layer, 1 Au layer and 2 dielectric (SiO2) spacer layers. First, we sputter a $\approx 3$~nm-thick film of WSi on a 75~mm silicon wafer with 150~nm of thermal oxide. The film was capped with $\approx 2$~nm $\alpha$Si to prevent oxidation. Next we patterned our wires using ZEP520A using electron-beam lithography. Then the film was etched using an inductively coupled plasma-reactive ion etcher (ICP-RIE) and an AR:SF6 etch. 

For the next layer, we pattern LOR3A + SPR660 using photolithography for the dielectric spacer layer. We sputter $\approx 25$~nm SiO2 and liftoff the resist---this layer isolates the WSi wire from the Nb rails. Then, we pattern LOR3A + SPR660 using photolithography for Nb liftoff. We then sputter $\approx 50$~nm of Nb and lift it off. Next, we pattern our rails using ZEP520A using electron-beam lithography. The Nb is etched using the ICP-RIE and an SF6 etch.

For the next layer, we pattern LOR3A + SPR660 using photolithography for the dielectric spacer layer. We sputter $\approx 25$~nm SiO2 and liftoff the resist---this layer isolates the Nb rails from the Au covers. Then we pattern taper covers in LOR3A + SPR660 using photolithography for Au liftoff. We evaporate $\approx 100$~nm Au and lift it off. 

\subsection{Verification of unity internal detection efficiency}\label{secA2}
As noted in Fig~2(a) in the main text, a slope on our detection plateau was evident for uncovered detectors of all widths. We hypothesize that this slope is a result of our experiments being performed under flood illumination causing detection events in the taper of our devices as has has been noted in previous studies~\cite{chiles2020superconducting}. To test this hypothesis, we fabricated several devices with gold covers placed over the taper in order to block light from hitting the tapers. In these gold covered devices, the only region expossed to light should be the uncovered portion of the constriction. In Fig~1(e) in the main text, we show an example of results obtained using a 50~$\mu$m device with a gold cover. The gold cover has a 10~$\mu$m-long opening over the center of the constriction. Here the slope appears to be completely eliminated by the addition of the cover. In the case of the 100~$\mu$m-wide SNSPDs with gold covers, we do not measure a complete reduction in plateau slope. Further study is needed to determine the cause of this discrepancy. 

We also verified unity internal detection efficiency using a focused beam spot on a 20~$\mu$m-wide SNSPD. Here we measured the photon count rate on the plateau as a function of rail current and found less than 1\% variation in the count rate over the range of rail current studied here.

\section{Field and current in the SNSPD-rail structure}\label{secS1}

The SNSPD is modeled by a long thin film wire of width $w$ situated at $y=0$ between two thin film rails of width $l$ at $w/2+b<x<w/2+b+l$ and $-w/2-b-l<x<-w/2-b$, as shown in Fig.~\ref{figD1}. Each rail is raised by the height $h$ and spaced by $b$ from the edge of the wire. It is assumed that the wire carries a dc current $I_s$ and each rail carries a dc current $I_r$, both the wire and the rails are infinite along $z$. 

\begin{figure}[h]
\centering
   	\includegraphics[scale=0.6,trim={0mm 0mm 0mm 0mm},clip]{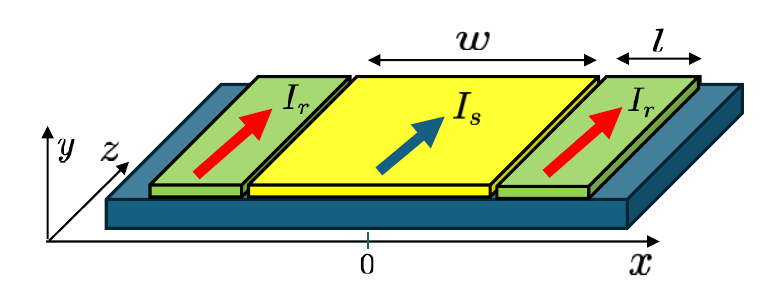}
   	\caption{Geometry of a detector film of width $w$ and two current-biased rails of width $l$. The coordinate axes are shifted downward for clarity.}
   	\label{figD1}
   \end{figure}

In this geometry the $z$-component of magnetic field $B_z$ along the wire vanishes and other field components $B_x=\partial A/\partial y$ and $B_y=-\partial A/\partial x$ are expressed in terms of the $z$ component of the vector potential ${\bf A}(x,y)=[0,0,A(x,y)]$ caused by currents flowing along $z$. Furthermore, $A(x,y)$ around the SNSPD in the thin film Pearl limit $d\ll\lambda$ in the absence of vortices satisfies the London equation:
\begin{equation}
\lambda^2\nabla^{2}A-dQ(x)\delta(y)=0,
\label{D1}
\end{equation}
where the gauge-invariant ${\bf Q}={\bf A}+(\phi_{0}/2\pi)\nabla\theta$ is proportional to the supercurrent density ${\bf j}=-{\bf Q}/\mu_0\lambda^2$ in the film, $\theta$ is the phase of the superconducting order parameter, $\phi_0$ is the magnetic flux quantum, $\lambda$ is the London penetration depth and $d$ is the film thickness. In the planar geometry considered here $j_x=j_y=0$, $j_z(x)$ is flowing along the wire length and the rail current can depend only on $x$, so the phase gradients $\nabla\theta=(0,0,\theta')$ have only a constant z-component $\theta'$ independent of $x$. 

The 2D London equation, Eq.~(\ref{D1}), was reduced to 1D integral equations for $Q_s(x)$ and $Q_r(x)$ in the inductively-coupled SNSPD and rails which then were solved numerically, as described in Ref.~\cite{ag}. Examples of the so-calculated $J(x)$ in wires of different widths are shown in Fig.~1(a) in the main text and Fig.~\ref{figD2}. Calculations were done for our WSi wires between thin film Nb rails with the respective Pearl lengths $\Lambda_s=2\lambda_s^2/d_s\approx 600\,\mu$m and $\Lambda_r=5\cdot 10^{-3}\Lambda_s$ for a moderately dirty Nb film. Particularly, Fig.~\ref{figD2}(b) shows that the rail-controlled reduction of $J(x)$ at the edges can be achieved in a wire 4 times wider than the Pearl length, which corresponds to a 2.4~mm-wide wire. The magnetic field around the SNSPD-rail structure is shown in Fig.~1(a) in the main text. 

\begin{figure}[h]
   	\includegraphics[scale=0.35,trim={5mm 0mm 0mm 0mm},clip]{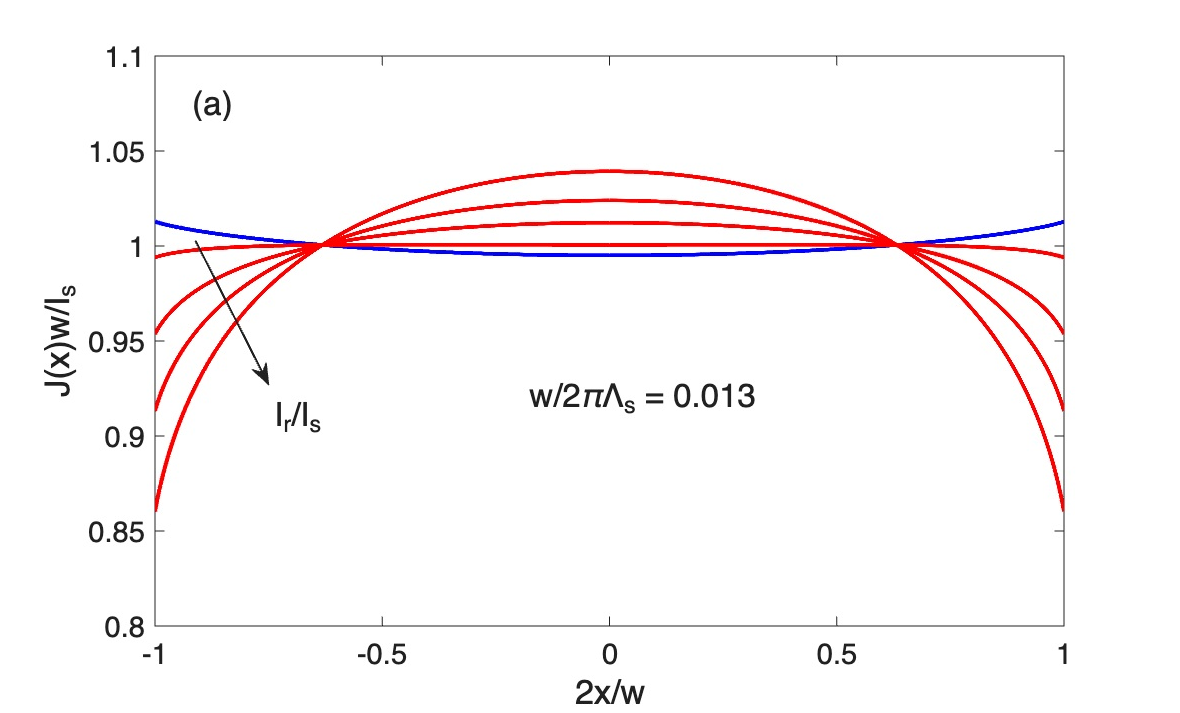}
	\includegraphics[scale=0.35,trim={10mm 0mm 0mm 0mm},clip]{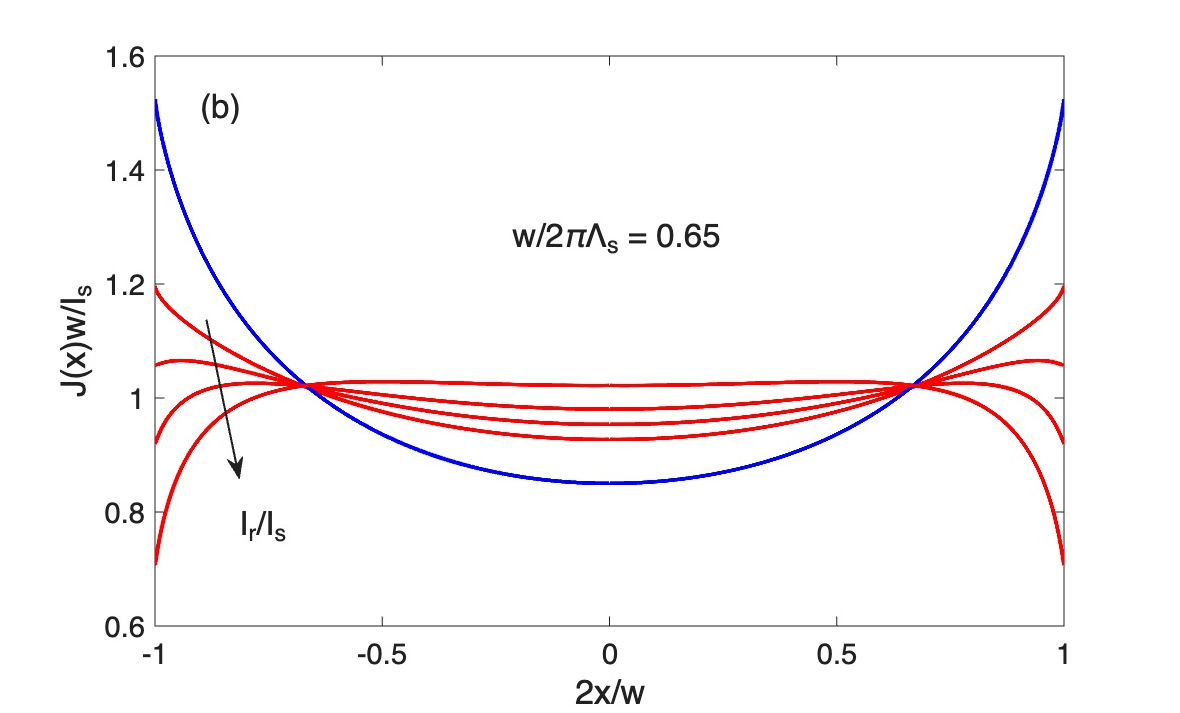}
   	\caption{Sheet current density $J(x)$ in the central wire calculated at different normalized rail currents $I_r/I_s$ for WSi films with $d_s=3$~nm, $\Lambda_s = 602\,\mu$m, $\Lambda_s=200\Lambda_r$, $l=0.15w$ and $b=0.005w$:  (a) $w\approx 50\,\mu$m, $I_r/I_s= 0.62, 1.89, 3.15,  4.81$ (b) $w\approx 2.5$~mm, $I_r/I_s=0.32, 0.41, 0.51, 0.65$.}
   	\label{figD2}
   	\end{figure}

\section{Energy barriers and dark count rates caused by vortices}\label{secS2}
Thermally-activated penetration of vortices through the film edges is determined by the position-dependent energy of a vortex $E(x)$ in a thin film, which has been calculated both in the London model~\cite{gurvink,ag} and the Ginzburg-Landau theory~\cite{Vodolaz}. At high operational current densities 
$J_s \sim J_d$ the energy barrier $U(J_s)=\mbox{max}[E(x)]$ is shifted very close to the film edge, as illustrated by Fig.~\ref{figE1}. In this case $U$ for a vortex in a narrow strip with ideal edges and $w\ll \Lambda_s$ is given by  
\begin{equation}
U\simeq\epsilon(1-J_s/J_d),
\label{E4}
\end{equation}
where $\epsilon=\phi_{0}^{2}/2\pi\mu_{0}\Lambda_s$ is the vortex line energy. For our 3~nm thick WSi films with $\lambda_s=950$~nm and $\Lambda_s\approx600\,\mu$m, we get $\epsilon\simeq 66$~K. At $J_s=0.9J_{d}$ and $T=0.9$~K the probability of thermally-activated penetration of a vortex from an ideal film edge is proportional to the Arrhenius factor $\propto\exp(-U/T)\sim 7\cdot10^{-4}$ which can increase substantially due to edge defects which locally reduce the energy barrier and facilitate penetration of vortices.
\begin{figure}[h]
\centering
   	\includegraphics[scale=0.4,trim={0mm 0mm 0mm 0mm},clip]{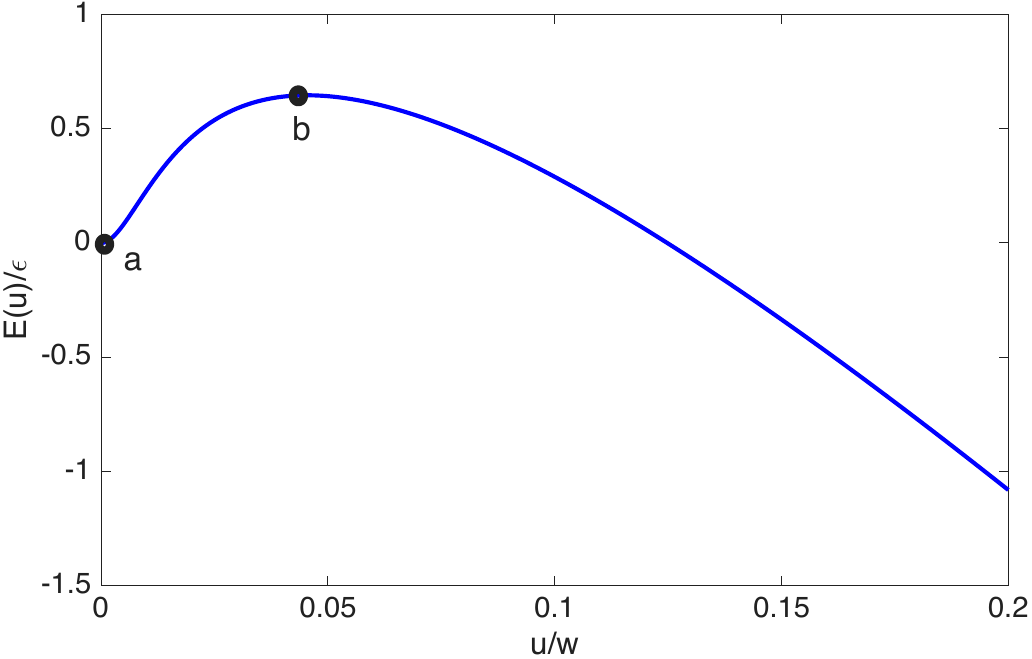}
   	\caption{Qualitative dependence of $E(x)$ on the vortex position evaluated from a regularized London model~\cite{ag} at $\xi=3\cdot 10^{-2}w$, $J_s=20J_0$ and $J_0=\phi_0/2\mu_0\Lambda_sw$ }
   	\label{figE1}
    \end{figure}
Edge defects can be small indentations or other lithographic features causing local variations of $T_c$ or current crowding around defects. For instance, the local current density at the edge of a semicircular indentation of radius $R\gg\xi$ is twice the mean $J_s$, and larger concentration of impurities at the edges causes local reduction of $J_d$. Such defects serve as gates for local penetration of vortices~\cite{det2} for which $U(J_s)$ vanishes at the mean current density $\bar{J}_d$ smaller than the bulk $J_d$. Given many material uncertainties depending on the film growth and deposition parameters, we assume that edge imperfections locally reduce the mean depairing current density to $\tilde{J}_d<J_d$ in a narrow strip along the edges. It is the reduction of $\tilde{J}_d$ at the edge which is mitigated by the current-biased rails. Next, consider the number $S_e$ of thermally-activated vortices penetrating over the edge barrier per unit time in a film of length $L$:
\begin{equation}
S_e = S_{e0}\exp\biggl[-\frac{\epsilon}{T}\left(1-\frac{J_e}{\tilde{J}_d}\right)\biggr],\qquad S_{e0}\sim \frac{L\nu}{l_i}.
\label{E5} 
\end{equation}
Here $J_e$ is the current density at the SNSPD edge controlled by $I_r$ in the rails, $L/\l_i$ is a number of statistically-independent vortex entries through edge defects with a mean spacing $l_i$ and $\nu$ is an attempt frequency. For the overdamped Abrikosov vortices, $\nu\simeq \sqrt{k_ak_b}/4\pi\eta$ follows from the classical result of Kramers~\cite{kramers}, where $k_a$ and $k_b$ are curvatures of $E(x)$ at the bottom and the top of the potential well depicted in Fig.~\ref{figE1}, $\eta=\phi_0^2d_s/2\pi\xi^{2}\rho_n$ is the Mattis-Bardeen vortex drag coefficient, and $\rho_n$ is the normal state resistivity. Taking $\sqrt{k_{a}k_{b}}\sim \phi_{0}^{2}d_s/4\pi\mu_{0}\lambda_s^2\xi_s^{2}$ at $J_s\sim \tilde{J}_d$ in Eq.~(\ref{E5}) yields
\begin{equation}
S_{e0}\sim \frac{L\rho_n}{8\pi\mu_0\lambda_s^2 l_i}.
\label{E6}
\end{equation}
For a WSi film with $L=0.1$~mm, $\lambda=950$~nm, $\rho_n=2.5\,\mu\Omega$m and $l_i\sim 10^2$ nm, we get $S_{e0}\sim 10^{12}$~s$^{-1}$. Here Eq.~(\ref{E6}) is a rough, order of magnitude estimate which takes into account neither microstructural details of edge imperfections nor deformation of the vortex core at the edge in the presence of current~\cite{Vodolaz}. The latter can result an extra factor $(1-J_s/J_d)^n$ with $n=1-2$, yet the logarithmic slope 
\begin{equation}
\frac{d\ln S_{e}}{dJ_s}=\frac{\epsilon}{T\tilde{J}_d} 
\label{E7}
\end{equation}
is practically insensitive to the uncertainties in $S_{e0}$ and its possible power-law dependence on $J_s$ if $T\ll\epsilon$.  

From Eq.~(\ref{E5}), we obtain a maximum current $I_{c1}$ which the SNSPD can carry at a given dark count rate $S_c$:
\begin{equation}
I_{c1}=\tilde{I}_d\left(1-\frac{T}{\epsilon}\ln\frac{S_{e0}}{S_c}\right).
\label{E8}
\end{equation} 
For $S_c=10^2$~s$^{-1}$ used in our definition of $I_{sw}$, $S_{e0}=10^{12}$~s$^{-1}$, $\epsilon = 66$~K, and $T = 0.9$~K, we get $I_{c1}\approx 0.69 \tilde{I}_d$, where $\tilde{I}_d=w\tilde{J}_d$. Such operational $I_{c1}$ equivalent to $I_{sw}$ is limited by the acceptable dark count rate $S_c$ and decreases logarithmically with the SNSPD length. The account of the extra factor $(1-J_s/\tilde{J}_d)$ in $S_{e0}$ only increases $I_{c1}$ by 3$\%$ and is neglected hereafter. 

\begin{figure}[h]
\centering
   	\includegraphics[scale=0.6,trim={0mm 0mm 0mm 0mm},clip]{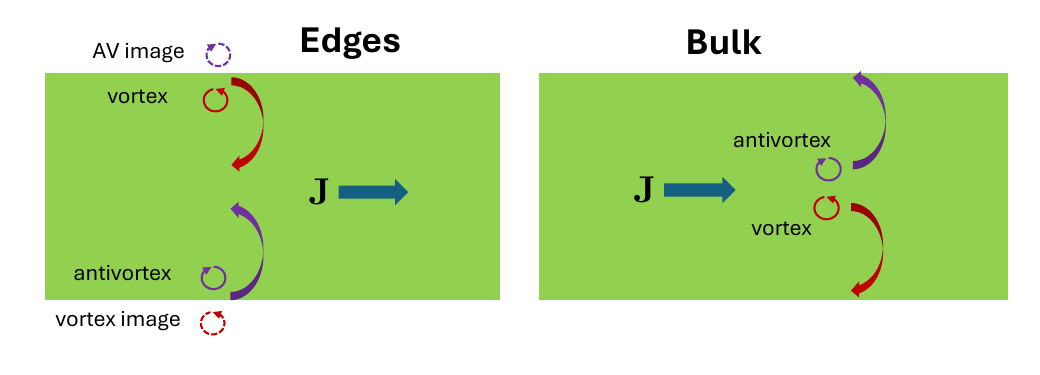}
   	\caption{Single vortices and antivortices penetrating from the film edges and vortex-antivortex pair unbinding inside the film due to thermal fluctuations.}
   	\label{figE2}
   	\end{figure}

Consider now the critical current $I_{c2}$ limited by thermally-activated unbinding of VAV pairs inside the film. This process is illustrated in Fig.~\ref{figE2}, which shows that the energy scale for the VAV pair production is twice that of the vortices penetrating from the edges: 
\begin{equation}
U=2\epsilon\left(1-\frac{J_s}{J_d}\right),\qquad J_s\sim J_d.
\label{E9}
\end{equation}
This barrier vanishes at the bulk $J_d$ larger than $\tilde{J}_c$ for penetration of vortices from the edges. The VAV production rate in the bulk of a film at $J_s\sim J_d$ can be evaluated in the same way as it was done above:
\begin{equation}
S_b=S_{b0}\exp\left[-\frac{2\epsilon}{T}\left(1-\frac{J_s}{J_d}\right)\right],\qquad S_{b0}\sim \frac{A\nu}{2\pi\xi_s^{2}}.
\label{E10}
\end{equation}
Here $A/2\pi\xi_s^2$ is a number of statistically-independent positions of the vortex core of size $\xi_s$ in a film of area $A$ and
\begin{equation}
S_{b0}\sim\frac{A\rho_n}{16\pi^2\mu_{0}\lambda_s^2\xi_s^{2}}. 
\label{E11}
\end{equation}
The critical current limited by the bulk VAV pair production is then
\begin{equation}
I_{c2}=I_{d}\left(1-\frac{T}{2\epsilon}\ln\frac{S_{b0}}{S_{c}}\right)
\label{E12}
\end{equation}
For $A=10\,\mu$m$\times 100\,\mu$m, $\rho_n=2.5\,\mu\Omega$m, $\lambda_s=950$~nm, $\xi_s=10$~nm, $\epsilon = 66$~K, and $T = 0.9$~K, Eqs.~(\ref{E11}) and (\ref{E12}) give $S_{b0}\sim 1\cdot10^{17}$~s$^{-1}$ and $I_{c2}\approx 0.76I_{d}$ at $S_c=10^2$~s$^{-1}$. Here $I_{c2}$ limited by the VAV unbinding decreases logarithmically with the SNSPD area.

As follows from Eqs.~(\ref{E7}) and (\ref{E12}), the logarithmic slopes of dark count rates for the VAV pair unbinding and the vortex hopping from the edges are related by:
\begin{equation}
\frac{d\ln S_b}{dI}=\left(\frac{2\tilde{J}_d}{J_d}\right)\frac{d\ln S_e}{d I}.
\label{E13}
\end{equation}
According to our experimental data shown in Fig.~1(d) in the main text, the slope $d\ln S/dI$ abruptly increases by a factor $1.79$ above a threshold current in the rails. This indicates switching from the dark counts limited by penetration of vortices from the edges to the bulk VAV pair unbinding. In this case Eq.~(\ref{E13}) can be used to evaluate the extent by which lithographic defects reduce the mean depairing current density at the edges: $\tilde{J}_d=0.895J_d$. Such relatively small decrease in $\tilde{J}_d$ is in line with the magnitude of reduction of $J(x)$ at the edges obtained in our numerical results shown in Fig.~\ref{figD2}(a).  

    \putbib[supplementary/Supbib]
  \end{bibunit}
\fi

\end{document}